\documentclass[twocolumn,showpacs,pra]{revtex4}
\usepackage{amssymb}
\begin{document}

\title{Lie fields revisited}
\author{Peter Morgan}
\email{peter.w.morgan@yale.edu}
\affiliation{Physics Department, Yale University, New Haven, CT 06520, USA.}
\homepage{http://pantheon.yale.edu/~PWM22}

\date{\today}
\begin{abstract}
A class of interacting classical random fields is constructed using deformed \mbox{$\star$-algebras}
of creation and annihilation operators.
The fields constructed are classical random field versions of ``Lie fields''.
A vacuum vector is used to construct linear forms over the algebras, which are conjectured to be states over
the algebras.
Assuming this conjecture is true, the fields constructed are ``quantum random fields'' in the sense that they
have Poincar\'e invariant vacua with a fluctuation scale determined by $\hbar$.
A nonlocal particle interpretation of the formalism is shown to be the same as a particle interpretation
of a quantum field theory.
\end{abstract}

\maketitle

\newcommand\Half{{\frac{1}{2}}}
\newcommand\Intd{{\mathrm{d}}}
\newcommand\RR {{\mathrm{I\hspace{-.1em}R}}}
\newcommand\CC{{{\rm C}\kern -0.5em 
          \vrule width 0.05em height 0.65em depth -0.03em
          \kern 0.45em}}
\newcommand\kT{{{\mathsf{k_B}} T}}
\def\Remove#1{{\raise 1.2ex\hbox{$\times$}\kern-0.8em \lower 0.35ex\hbox{$#1$}}}
\newcommand\eqdef{{\stackrel{\mathrm{def}}{=}}}
\newcommand\DP{{\diamond\kern-0.65em +}}
\newcommand{\SmallFrac}[2]{{\scriptstyle\frac{\scriptstyle #1}{\scriptstyle#2}}}
\newcommand{\BLow}[1]{{\lower 0.65ex\hbox{${}_{#1}\!\!$}}}

\section{Introduction}
Instead of the interacting Hamiltonian and Lagrangian methods that require regularization and renormalization,
we will here pursue a purely algebraic approach, which will lead to a deformation of the commutation relations
of an algebraic presentation of the generalized free classical random field.
The classical random field models constructed here are based on the structure of the quantum
fields known as ``Lie fields''\cite{GreenbergA,Lowenstein,GreenbergB,Baumann}.
Lowenstein\cite[p. 57]{Lowenstein} points out that the examples he constructs do not lead to scattering,
which Greenberg\cite{GreenbergB} proves for quantum Lie fields in general.
Baumann\cite{Baumann} proves that Lie fields are incompatible with the spectrum condition, but classicality
of a \mbox{$\star$-algebra} of observables is a different constraint than microcausality and the spectrum condition,
so that we will be able to construct ``classical random Lie fields with quantum fluctuations''.

A previous paper presented the classical Klein-Gordon random field using a creation and annihilation operator
formalism, and showed that in these terms quantum fluctuations and thermal fluctuations can be presented
as Poincar\'e invariant and non-Poincar\'e invariant fluctuations respectively\cite{MorganA}.
It was also shown that we can construct a Poincar\'e invariant state of the classical Klein-Gordon random
field, with a fluctuation scale determined by $\hbar$, that produces measurement results identical
to measurement results produced by the vacuum state of the quantized Klein-Gordon field, whenever
measurements are made at space-like separation.
When incompatible measurements are made at time-like separation, so that measurements affect the results
of subsequent measurements, careful attention to the differences between quantum and classical models of
measurement is needed\cite{MorganB}.
Classical models must explicitly model the effects of quantum fluctuations on measurement by including
the measurement apparatus in the model whenever the effects of quantum fluctuations are empirically significant.
Appendix \ref{MeasurementTheory} discusses the relationship between the measurement theories of quantum
fields and of classical random fields.

Section \ref{freeKG} of this paper discusses the free Klein-Gordon quantum field and the free Klein-Gordon
classical random field using a creation and annihilation operator formalism, then section \ref{intKG} constructs
scalar classical random Lie fields.
A vacuum vector, defined as a zero eigenvector of all annihilation operators, is used to construct a linear form
over the algebra of creation and annihilation operators.
A partial proof that this linear form is a state over the algebra is given in Appendix \ref{VacuumState},
but part of the proof remains a conjecture.
In corroboration of the conjecture, Appendix \ref{VacuumState} shows that the linear form is positive
semi-definite for operators that are constructed as sums of products of up to five annihilation operators
and five creation operators.

Section \ref{MeasurementAndScattering} discusses measurement and scattering theory for these Lie fields, which
gives a sense in which a particle interpretation of the formalism is the same as a particle interpretation of a
quantum field theory, even though as a local operator model the Lie fields we have constructed are certainly not
local quantum field theories.
Section \ref{nonScalarLieFields} constructs Lie fields based on vector or electromagnetic fields.

Statistics of observables and correlations between them are a principle empirical foundation of modern physics.
This paper constructs a continuous range of deformations of generalized free fields as algebraic models for
those statistics and correlations that goes beyond the range of application of free fields.

I gratefully acknowledge the contributions of a very helpful referee.

\section{The free Klein-Gordon field and the generalized free field}\label{freeKG}
Measurement for a classical Klein-Gordon random field can be modelled by an operator $\hat\phi_f$ that
satisfies $[\hat\phi_f,\hat\phi_g]=0$ for all $f$ and $g$ in a test function space $\mathcal{S}$ on
Minkowski space, which we will take to be a Schwartz space of complex-valued functions\cite[\S II.1.2]{Haag},
so that $f(x)$ is infinitely often differentiable and decreases as well as its derivatives faster than any
power as $x$ moves to infinity in any direction.
Although $\hat\phi_f$ can be presented in terms of an operator-valued distribution $\hat\phi(x)$ as
$\int \Intd^4x f(x)\hat\phi(x)$, we will work here mostly with operators indexed by test functions,
which gives an effective and compact notation.
We may construct the classical Klein-Gordon random field as a sum of creation and annihilation operators,
$\hat\phi_f=a_f+a_{f^*}^\dagger$\cite{MorganA}.
The construction $\hat\phi_f=a_f+a_{f^*}^\dagger$ makes $\hat\phi_f$ complex linear in $f$, supposing that
we define $a_f$ to be complex linear in $f$ --- this is essential for a Wightman field and mathematically very
convenient, but $\hat\phi_f$ is an observable only when $f=f^*$ is real, so that $\hat\phi_f^\dagger=\hat\phi_f$.
An equilibrium state of the classical Klein-Gordon random field at temperature $T$ can be presented in terms
of a Euclidean invariant ``vacuum'' state $\left|0\right>$, for which $a_f\left|0\right>=0$, if $a_f$ and
$a_g^\dagger$ satisfy the commutation relations $[a_f,a_g^\dagger]=(g,f)_C$ and $[a_f,a_g]=0$.
The non-Lorentz invariant inner product $(g,f)_C$ is given by
\begin{equation}\label{GibbsState}
  (g,f)_C=\kT \int \frac{\Intd^4k}{(2\pi)^4}\,\frac{2\pi\delta(k^\mu k_\mu-m^2)}{k_0}\tilde g^*(k)\tilde f(k).
\end{equation}
This contrasts with the inner product that determines the structure of the quantized Klein-Gordon field,
\begin{equation}
  (g,f)_{Q+}=\hbar \int \frac{\Intd^4k}{(2\pi)^4}\,2\pi\delta(k^\mu k_\mu-m^2)\theta(k_0)\tilde g^*(k)\tilde f(k),
\end{equation}
which explicitly restricts to the forward light-cone as well as being Lorentz invariant because of the
removal of the $k_0$ from the denominator, resulting in the Poincar\'e invariant vacuum state of the
quantized Klein-Gordon field.
Given the vacuum state over the \mbox{$\star$-algebra} generated by the creation and annihilation operators, we can
use the GNS construction to construct a Hilbert space $\mathcal{H}_C$\,\cite[\S III.2.2]{Haag}.
The GNS representation of the classical unbounded random field operator $\hat\phi_f$ acts on a dense domain
$\mathcal{D}_C\subset\mathcal{H}_C$.

In a creation and annihilation operator formalism, it is the restriction to the forward light-cone, the
spectrum condition, that results in nontrivial commutation relations $[\hat\phi_f,\hat\phi_g]$ when the
supports of $f$ and $g$ are not at space-like separation.
We can also construct a Poincar\'e invariant state of the classical Klein-Gordon random field, in which
$k_0$ is removed from the denominator, but no restriction to the forward light-cone is imposed,
\begin{equation}
  (g,f)_{QC}=\frac{\hbar}{2}
                  \int \frac{\Intd^4k}{(2\pi)^4}\,2\pi\delta(k^\mu k_\mu-m^2)\tilde g^*(k)\tilde f(k).
\end{equation}
This produces measurement results that are identical to measurement results produced by the vacuum state of
the quantized Klein-Gordon field, whenever measurements are made at space-like separation.
Measurements can be made at time-like separation in this classical random field model, but are incompatible
in a quantum field model, so a direct comparison of time-like separated measurements is not possible.
This Poincar\'e invariant state does not maximize the entropy of the Klein-Gordon random field, so we can
either introduce a classical dynamics with the nonlocal Hamiltonian
$\int \Intd^3\mathbf{k}\tilde\Phi^*(\mathbf{k})\sqrt{|\mathbf{k}|^2+m^2}\tilde\Phi(\mathbf{k})$, which
has this Poincar\'e invariant state as its Gibbs equilibrium state, or else we can
introduce a ``Poincar\'e invariant entropy'', which we might call ``quantum entropy'', as the
thermodynamic dual to $\hbar$ (this is in addition to entropy, which is the thermodynamic dual to
temperature --- in \cite{MorganA} I preferred the introduction of a nonlocal dynamics, but the
idea of quantum, Poincar\'e invariant entropy has interesting consequences, to be developed elsewhere).

Note that on a view of the quantized Klein-Gordon field that is suggested by comparison with a classical
random field, the spectrum condition has nothing to do with ``stability'', which is the reason generally given
for insisting on the restriction to the forward light-cone.
For the classical model presented by equation (\ref{GibbsState}), it is the maximization of the entropy,
and hence the minimization of the classical free energy, that makes the Gibbs state stable --- the classical
energy of the Gibbs state is infinite, but lower temperature states are thermodynamically inaccessible.

The generalized free quantum field smears the mass shell by a K\"all\'en-Lehmann weight function
$\rho(\sigma)$, 
\begin{equation}
  (g,f)_{gQ+}=\hbar \int \frac{\Intd^4k}{(2\pi)^4}\Intd\sigma
       \rho(\sigma)2\pi\delta(k^\mu k_\mu-\sigma)\theta(k_0)\tilde g^*(k)\tilde f(k).
\end{equation}
See Streater\cite[\S 3.4]{Streater} for a brief review of properties of the generalized free quantum field.

\section{Interacting creation and annihilation operators for a scalar field}\label{intKG}
We will consider here deformations of the commutation relations for creation and annihilation relations
that satisfy linearity, $\lambda a_f+\mu a_g=a_{\lambda f+\mu g}$.
Using creation and annihilation operators makes possible the construction of a classical
\mbox{$\star$-algebra} that parallels the construction of a quantum Lie field\cite{Lowenstein}.
Linearity is a very tight constraint, insofar as the only choice for a deformed commutation relation is
\begin{equation}\label{CommRel}
  [a_f,a_g^\dagger]=(g;f)+a_{\xi(g;f)}+a_{\xi(f;g)}^\dagger,
\end{equation}
where $\xi(g;f)$ is test function valued and anti-linear and linear in the test functions $g$ and $f$ respectively.
We will suppose that the inner product $(\cdot;\cdot)$ is Lorentz invariant, but may be either
quantum or classical in the sense of the previous section, restricting the energy-momentum spectrum either
to the forward light-cone or to the forward and backward light-cones.
The only other possible linear deformation, $q$-deformation of the commutation relation, is ruled out
because it does not allow either microcausality or classicality to be satisfied, and we will not consider
here the possibility of deformed anti-commutation relations.
Non-scalar fields are constructed in section \ref{nonScalarLieFields}.
The literature on quantum Lie fields requires that $[\hat\phi_f,\hat\phi_g]=i\omega(f,g)+\hat\phi_{\xi(f,g)}$,
which requires a commutation relation of the form (\ref{CommRel}) for creation and annihilation operators.
For a classical random field, the commutation relation $[\hat\phi_f,\hat\phi_g]=0$ is trivial, so we will
define the algebraic structure of a ``classical random Lie field'' by equation (\ref{CommRel}).

We can always apply the commutation relations (\ref{CommRel}) to convert any product of creation and
annihilation operators to a sum of normal-ordered products
$a_{g_1}^\dagger...a_{g_m}^\dagger a_{f_1}...a_{f_n}$, so that we can construct a \mbox{$\star$-algebra} that
is generated by a finite but arbitrarily large set of test functions, presuming that associativity
can be ensured for some choice of $(\cdot;\cdot)$ and $\xi(\cdot;\cdot)$.
To ensure associativity of the \mbox{$\star$-algebra} generated by these creation and annihilation operators, we must
ensure that the Jacobi identity is satisfied for all elements in the algebra; as a first step, we must
ensure that single creation and annihilation operators satisfy the Jacobi identity,
\begin{equation}
  [a_f,[a_g,a_h^\dagger]]+[a_g,[a_h^\dagger,a_f]]+[a_h^\dagger,[a_f,a_g]]=0,
\end{equation}
which requires that
\begin{eqnarray}
  &&(\xi(g;h);f)=(\xi(f;h);g),        \label{C1}\\
  &&\xi(f;\xi(g;h))=\xi(g;\xi(f;h)),  \label{C2}\\
  &&\xi(\xi(g;h);f)=\xi(\xi(f;h);g).  \label{C3}
\end{eqnarray}
If we can find functions that satisfy equations (\ref{C1}), (\ref{C2}), and (\ref{C3}), then, because
of the symmetries of $(\cdot;\cdot)$ and $\xi(\cdot;\cdot)$, we can define two more general objects,
which take two lists of functions as arguments, symmetric in their anti-linear and linear
arguments separately,
\begin{eqnarray*}
  \xi(\xi(f_1,...,f_m;g_1,...,g_n);f)&=& \xi(g_1,...,g_n;f_1,...,f_m,f) \cr
  \xi(g;\xi(g_1,...,g_n;f_1,...,f_m))&=& \xi(g_1,...,g_n,g;f_1,...,f_m) \cr
  (\xi(f_1,...,f_m;g_1,...,g_n);f)   &=& (g_1,...,g_n;f_1,...,f_m,f) \cr
  (g;\xi(g_1,...,g_n;f_1,...,f_m))   &=& (g_1,...,g_n,g;f_1,...,f_m).
\end{eqnarray*}
For example,
\begin{eqnarray*}
  (\xi(f_1;g_1);f_2)&=&(g_1;f_1,f_2),\cr
  \xi(\xi(f_1;g_1);\xi(g_2,f_2))&=&\xi(g_1,g_2;f_1,f_2),\cr
  (\xi(f_1;\xi(f_3;g_1));\xi(\xi(f_4;g_2);f_2))&=&(g_1,g_2;f_1,f_2,f_3,f_4).
\end{eqnarray*}
The position of each function in the flattened expression depends only on whether it is linear or
antilinear in the original expression.
This can be extended by linearity to a sesquilinear form on the symmetrized tensor product spaces
$S(\mathcal{S}^{\otimes n})$ over the test function space, for $n>1$.
With this definition, we can write the product $a_g a^\dagger_{f_1}\cdots a^\dagger_{f_n}$ as
\begin{widetext}
\begin{eqnarray}\label{productn1}
%
  a_g a^\dagger_{f_1}\cdots a^\dagger_{f_n} &=& a^\dagger_{f_1}\cdots a^\dagger_{f_n}a_g+  
     \sum_{i=1}^n a^\dagger_{f_1}\cdots \Remove{a^\dagger_{f_i}}\cdots a^\dagger_{f_n}
              \left((f_i;g)+a^\dagger_{\xi(g;f_i)}+a_{\xi(f_i;g)}\right)+\cr
   &&\quad \sum_{i=1}^n\sum_{j=i+1}^n
              a^\dagger_{f_1}\cdots \Remove{a^\dagger_{f_i}}\cdots \Remove{a^\dagger_{f_j}}\cdots a^\dagger_{f_n}
              \left((f_i,f_j;g)+a^\dagger_{\xi(g;f_i,f_j)}+a_{\xi(f_i,f_j;g)}\right)+
   \quad ...\quad + \cr\vspace{-1ex}\cr
   &&\quad \left((f_1, f_2, ..., f_n;g)+a^\dagger_{\xi(g;f_1, f_2, ...,f_n)}+a_{\xi(f_1, f_2, ..., f_n;g)}\right)\cr
   &&\hspace{-5em}
     =\sum_{\sigma\in S_n}\sum_{k=0}^{n-1}\frac{1}{k!(n-k)!}\left[\prod_{j=1}^k a^\dagger_{f_{\sigma(j)}}\right]
      \left( (f_{\sigma(k+1)},...,f_{\sigma(n)};g)+a^\dagger_{\xi(g;f_{\sigma(k+1)},...,f_{\sigma(n)})} +
                                                   a_{\xi(f_{\sigma(k+1)},...,f_{\sigma(n)};g)}\right),
\end{eqnarray}
\end{widetext}
where $\Remove{a^\dagger_{f_i}}$ denotes that this entry in a list is removed.
This product of one annihilation operator with $n$ creation operators (which generates $3\cdot 2^n-2$
terms, in contrast to the $n+1$ terms that are generated by the same product of undeformed annihilation
and creation operators) is symmetric in $f_1,...,f_n$.
The Jacobi identity is satisfied for general basis elements of this \mbox{$\star$-algebra} because of the symmetry
of equation (\ref{productn1}) in $f_1,...,f_n$, and similar symmetries of higher products, so that we
obtain the same final expression in terms of normal-ordered expressions in the annihilation and creation
operators independently of the order in which we apply the commutation relation (\ref{CommRel}) to move
creation operators to the left and annihilation operators to the right.

We will here assume that the linear form generated by the vacuum vector, $\omega(\hat A)=\left<0\right|\hat A\left|0\right>$
is a state over the \mbox{$\star$-algebra} generated by creation and annihilation operators that satisfy the commutation
relation (\ref{CommRel}), as it is for the free field, so that $\left<0\right|\hat A\hat A^\dagger\left|0\right> \ge 0$
for all operators in the \mbox{$\star$-algebra}.
Appendix \ref{VacuumState} gives a partial proof and a substantial corroboration that this is true for a classical
random Lie field.
We can straightforwardly compute the lowest degree moments of the probability density associated with the
observable $\hat\phi_f=a_f+a_{f^*}^\dagger$, in the state $\left<0\right|\hat A\left|0\right>$ that is
defined by $a_f\left|0\right>=0$,
\begin{eqnarray}
  \left<0\right|\hat\phi_f\left|0\right> &=& 0\cr
  \left<0\right|\hat\phi_f^2\left|0\right> &=& (f^*;f)\cr
  \left<0\right|\hat\phi_f^3\left|0\right> &=& (f^*;f,f)+(f^*,f^*;f)\cr
  \left<0\right|\hat\phi_f^4\left|0\right> &=& (f^*;f,f,f)+4(f^*,f^*;f,f)+\cr
                                            && \quad(f^*,f^*,f^*;f)+3(f^*,f)^2.
\end{eqnarray}
The expressions for higher moments quickly become more complex, but the cumulants of the probability
density are straightforward,
\begin{eqnarray}
  C_1(f) &=& 0\cr
  C_2(f) &=& (f^*;f)\cr
  C_3(f) &=& (f^*;f,f)+(f^*,f^*;f)\cr
  C_4(f) &=& (f^*;f,f,f)+4(f^*,f^*;f,f)+(f^*,f^*,f^*;f)\cr
  C_5(f) &=& (f^*;f,f,f,f)+11(f^*,f^*;f,f,f)+\cr
          && \quad 11(f^*,f^*,f^*;f,f)+(f^*,f^*,f^*,f^*;f)\cr
  ...\cr
  C_n(f) &=& \sum_{k=1}^{n-1} \left<{n-1\atop k-1}\right>\,(f^{*\,\times k};f^{\times (n-k)}),
\end{eqnarray}
where $\left<{n\atop k}\right>$ are Eulerian numbers, which satisfy the recurrence relation
\begin{displaymath}
  \left<{n\atop k}\right>=(n-k)\left<{n-1\atop k-1}\right>+(k+1)\left<{n-1\atop k}\right>.
\end{displaymath}
We can also calculate $n$-measurement connected correlation functions, obtaining for the 3- and
4-measurement connected correlation functions, for example,
\begin{eqnarray*}
  \left<0\right|\hat\phi_{f_1}\hat\phi_{f_2}\hat\phi_{f_3}\left|0\right>_c &=& (f_3^*;f_2,f_1)+(f_3^*,f_2^*;f_1)\cr
  \left<0\right|\hat\phi_{f_1}\hat\phi_{f_2}\hat\phi_{f_3}\hat\phi_{f_4}\left|0\right>_c&=&
          (f_4^*;f_3,f_2,f_1)+3(f_4^*,f_3^*;f_2,f_1)+\cr
          && \quad(f_4^*,f_2^*;f_3,f_1)+(f_4^*,f_3^*,f_2^*;f_1).
\end{eqnarray*}
As an aside, the $n$-measurement connected correlated functions do not vanish at any order, so the conditions
for the proof of Greenberg and Licht for Wightman theories\cite{GL}, that there is no scattering if the
truncated functions vanish beyond some order, are not satisfied.

To find a space-time model for $(\cdot;\cdot):\mathcal{S}\times\mathcal{S}\rightarrow \RR$ and
$\xi(\cdot;\cdot):\mathcal{S}\times\mathcal{S}\rightarrow\mathcal{S}$, we will use the following
momentum space expressions,
\begin{eqnarray}
  (g;f)&=&\int\Intd^4u M(u)\tilde g^*(u)\tilde f(u),  \label{A1}\\
  \tilde \xi(g;f)(s)&=&\int\Intd^4u\Intd^4v M(u)\tilde g^*(u)\tilde f(v)\delta(s+u-v)F(s,u)  \cr
                    &=&\int\Intd^4u M(u)\tilde g^*(u)\tilde f(u+s)F(s,u),  \label{A2}
\end{eqnarray}
where we will consider cases in which the mass function $M(u)$ may be quantum (non-zero only on and
within the forward light-cone) or classical (non-zero on and within both the forward and backward
light-cones).
The delta function $\delta(s+u-v)$ immediately ensures translation invariance, for which we must have
$(g_x;f_x)=(g;f)$ and $\xi_x(g_x;f_x)=\xi(g;f)$, where $g_x(y)=g(y-x)$.
Equation (\ref{C1}) then requires that
\begin{widetext}
\begin{eqnarray}
  (f;\xi(g;h))&=&\int\Intd^4u M(u)\tilde f^*(u)\tilde \xi(g,h)(u) \cr
      &=&\int\Intd^4u\Intd^4v M(u)\tilde f^*(u) M(v)\tilde g^*(v)\tilde h(u+v) F(u,v)
\end{eqnarray}
must be equal to $(g;\xi(f;h))$, which is satisfied if $F(u,v)=F(v,u)$ is symmetric under
permutation of $u,v$ for almost all $u, v$.
Equation (\ref{C2}) requires that
\begin{eqnarray}
  \tilde \xi(f;\xi(g;h))(s)&=&\int\Intd^4u M(u)\tilde f^*(u)\tilde \xi(g,h)(s+u)F(s,u)\cr
      &=&\int\Intd^4u\Intd^4v M(u)\tilde f^*(u)M(v)\tilde g^*(v)\tilde h(s+u+v)F(s+u,v)F(s,u)\cr&&
\end{eqnarray}
must be equal to $\tilde \xi(g;\xi(f;h))(s)$, which is satisfied if 
\begin{equation}\label{ConditionB}
  F(s+u,v)F(s,u)=F(s+v,u)F(s,v)
\end{equation}
is symmetric under permutation of $u,v$ for almost all $u, v$.
Finally, equation (\ref{C3}) requires that
\begin{eqnarray}
  \tilde \xi(\xi(g;h);f)(s)&=&\int\Intd^4u M(u)\tilde \xi^*(g,h)(u)\tilde f(s+u)F(s,u)  \cr
       &=&\int\Intd^4u\Intd^4v \tilde g(v) \tilde h^*(u+v)\tilde f(s+u)F^*(u,v)F(s,u)M(u)M(v)  \cr
       &=&\int\Intd^4u\Intd^4v' \tilde g(v'+s)\tilde f(u+s) \tilde h^*(u+v'+s)F^*(u,v'+s)F(s,u)M(u)M(v'+s)\cr&&
\end{eqnarray}
must be equal to $\tilde \xi(\xi(f;h);g)(s)$, which is satisfied if
\begin{equation}\label{ConditionC}
  F^*(u,v'+s)F(s,u)M(u)M(v'+s)=F^*(v',u+s)F(s,v')M(v')M(u+s)\ 
\end{equation}
\end{widetext}
is symmetric under permutation of $u,v'$ for almost all $u, v'$.
These three symmetries can be solved, provided $F(s,u)$ and $M(u)$ are almost always non-zero,
by multiplying equation (\ref{ConditionC}) by $F^*(s,u)$ and using equation (\ref{ConditionB})
and the symmetry $F(s,u)=F(u,s)$, so that we obtain
\begin{equation}
  \frac{|F(s,u)|^2 M(u)}{M(u+s)}=\frac{|F(s,v)|^2 M(v)}{M(v+s)}, 
\end{equation}
which, taking into account again the symmetry $F(s,u)=F(u,s)$, requires
\begin{equation}
  |F(s,u)|^2=\lambda^2\frac{M(u+s)}{M(u)M(s)}.
\end{equation}
With this solution, equations (\ref{A1}) and (\ref{A2}) become
\vspace{0.1pt plus 2ex}
\begin{eqnarray}
  (g;f)&=&\int\Intd^4u\; \tilde g^*(u)\tilde G^*(u)\tilde G(u)\tilde f(u),\\
  \tilde \xi(g;f)(s)&=&\lambda\int\Intd^4u\Intd^4v\;\tilde g^*(u)\frac{\tilde G^*(u)\tilde G(v)}{\tilde G(s)}\tilde f(v)
                              \delta(s+u-v)\cr
                    &=&\lambda\int\Intd^4u\;\tilde g^*(u)\frac{\tilde G^*(u)\tilde G(u+s)}{\tilde G(s)}\tilde f(u+s),
\end{eqnarray}
where $|\tilde G(u)|^2=M(u)$.
Using the above, we can construct the sesquilinear forms $(...;...)$ and $\xi(...;...)$,
\begin{widetext}
\begin{eqnarray}
    (g_1,...,g_m;f_1,...,f_n)&=&\lambda^{m+n-2}
                               \int\delta\left(\sum_{i=1}^m u_i-\sum_{j=1}^n v_j\right)
                               \prod_{i=i}^m\Intd^4u_i \tilde G^*(u_i)\tilde g_i^*(u_i)
                               \prod_{j=1}^n\Intd^4v_j \tilde G(v_j)\tilde f_j(v_j),  \label{A1M}\\
  \tilde \xi(g_1,...g_m;f_1,...,f_n)(s)&=&\frac{\lambda^{m+n-1}}{\tilde G(s)}
                               \int\delta\left(s+\sum_{i=1}^m u_i-\sum_{j=1}^n v_j\right)
                               \prod_{i=i}^m\Intd^4u_i \tilde G^*(u_i)\tilde g_i^*(u_i)
                               \prod_{j=1}^n\Intd^4v_j \tilde G(v_j)\tilde f_j(v_j).  \label{A2M}
\end{eqnarray}
\end{widetext}
The restriction that $M(u)\ne 0$ almost everywhere is not empirically significant, insofar as $M(u)$
may be arbitrarily close to zero, but also note that $(...;...)$ does not require
the inverse of the mass function at all.
$\xi(...;...)$ does not appear in the final expressions for expected values of observables, except
implicitly as part of the construction of $(...;...)$, so for all observables we
can comfortably take the limit in which $M(u)\rightarrow 0$ outside the light-cone.
For the classical random field we will construct, we can modify the algebra so that we may take $M(u)=0$
on any Lorentz-invariant set, in particular so that $M(u)=0$ for every space-like $u$, without having to
take limits (see Eq. (\ref{gimelCommRel})\hspace{0.1em}).
For the deformed commutation algebra to exist, note that $|\tilde G(u)|^2=M(u)$, so $M(u)$ cannot be restricted
to a single mass shell as a distribution, we must take as a starting point a generalized free field that
has a K\"all\'en-Lehmann weight function of which we can legitimately take a square root.

Having constructed the associative \mbox{$\star$-algebra} above, we must now consider the possibility of microcausality
or classicality for the observables $\hat\phi_f=a_f+a^\dagger_{f^*}$, $[\hat\phi_f,\hat\phi_g]=0$ either for
functions that have space-like separated supports or for all functions, respectively.
\begin{equation}
  [\hat\phi_f,\hat\phi_g]=(g^*;f)-(f^*;g)+a_{\xi(g^*;f)-\xi(f^*;g)}+a_{\xi(f;g^*)-\xi(g;f^*)}^\dagger
\end{equation}
will be zero whenever
\begin{equation}\label{MCcondition}
  (g^*;f)=(f^*;g)\ \mathrm{and}\ \xi(g^*;f)=\xi(f^*;g)
\end{equation}
are both symmetric in $f$ and $g$.
The first condition is satisfied for a generalized free quantum field, when $f$ and $g$ have space-like
separated supports.
The first condition is also satisfied for a generalized free classical random field, for all test functions
$f$ and $g$.
The symmetry condition $\xi(g^*;f)=\xi(f^*;g)$ can be satisfied for test functions $f$ and $g$, noting that
$\widetilde{g^*}(u)=\tilde g^*(-u)$,
\begin{widetext}
\begin{eqnarray}
  \tilde \xi(g^*;f)(s)&=&\lambda\int\Intd^4u\Intd^4v\;\tilde g(-u)\frac{\tilde G^*(u)\tilde G(v)}{\tilde G(s)}\tilde f(v)
                        \delta(s+u-v)\cr
     &=&\lambda\int\Intd^4u\Intd^4v\;\tilde g(u)\frac{\tilde G^*(-u)\tilde G(v)}{\tilde G(s)}\tilde f(v)\delta(s-u-v),
\end{eqnarray}
\end{widetext}
only if $\tilde G^*(-u)=\tilde G(u)$ for almost all $u$, which can be satisfied for a classical random field, but is
incompatible with the spectrum condition\cite{Baumann}.

In the classical case, with which we will be solely concerned from now on, $\tilde G^*(-u)=\tilde G(u)$ is satisfied
in a Poincar\'e invariant way by \mbox{$\tilde G(u)=G_{0}(u_\mu u^\mu)e^{iG_{1}(u_\mu u^\mu)(\theta(u_0)-\theta(-u_0))}$},
for arbitrary real functions $G_{0}$ and for all real functions $G_{1}$ that are zero for space-like $u_\mu$.
The classical sesquilinear form $(g_1,...,g_n;f_1,...,f_m)$ is equal to 
$(g_1,...,g_{n-1};g_n^*,f_1,...,f_m)$, \emph{etc.}, so we can write this as
$(;g_1^*,...,g_n^*,f_1,...,f_m)$, and similarly $\xi(g_1...,g_n;f_1,...,f_m)$ can be rewritten as
$\xi(g_1,...,g_{n-1};g_n^*,f_1,...,f_m)$, \emph{etc.}, so the $n$-measurement connected correlation
function $\left<0\right|\hat\phi_{f_1}\cdots\hat\phi_{f_n}\left|0\right>_c$ reduces for $n>1$ to
$(n-1)!(;f_1,...,f_n)$; for $n=1$, $\left<0\right|\hat\phi_{f_1}\left|0\right>=0$.
Defining a notation $\tilde f^\gimel(u)=\tilde G(u)\tilde f(u)$, of which the inverse Fourier transform $f^\gimel(y)$
can be constructed because $f(x)$, as a Schwartz space function, is well-behaved, then
$(;f_1,...,f_n)$ can be straightforwardly evaluated in real-space as
\begin{equation}
(;f_1,...,f_n)=\lambda^{n-2}(2\pi)^{4(n-1)}\int f_1^\gimel(y)\cdots f_n^\gimel(y)\Intd^4y,
\end{equation}
so the $n$-measurement connected vacuum correlation function is a measure of the ``overlap'' of the
Lorentz-invariantly $G$-modified smearing functions $f_1^\gimel$, $f_2^\gimel$, ..., $f_n^\gimel$.
For example,
\begin{widetext}
\begin{eqnarray}
  (;f_1,f_2,f_3)&=&\lambda\int\tilde f_1^\gimel(u_1)\tilde f_2^\gimel(u_2)\tilde f_3^\gimel(u_3)
                                    \delta(u_1+u_2+u_3)\Intd^4u_1\Intd^4u_2\Intd^4u_3\cr
                &=&\lambda\int f_1^\gimel(x_1)e^{iu_1x_1}f_2^\gimel(x_2)e^{iu_2x_2}f_3^\gimel(x_3)e^{iu_3x_3}
                                    \delta(u_1+u_2+u_3)\Intd^4u_1\Intd^4u_2\Intd^4u_3\Intd^4x_1\Intd^4x_2\Intd^4x_3\cr
                &=&\lambda\int f_1^\gimel(x_1)e^{iu_1(x_1-x_3)}f_2^\gimel(x_2)e^{iu_2(x_2-x_3)}f_3^\gimel(x_3)
                                    \Intd^4u_1\Intd^4u_2\Intd^4x_1\Intd^4x_2\Intd^4x_3\cr
                &=&\lambda(2\pi)^8\int f_1^\gimel(x_3)f_2^\gimel(x_3)f_3^\gimel(x_3)\Intd^4x_3
\end{eqnarray}
\end{widetext}
We can also compactly express $\xi^\gimel(;f_1,...,f_n)$ as a real-space product,
\begin{equation}\label{XiGimel}
  \xi^\gimel(;f_1,...,f_n)(y)=\lambda^{n-1}(2\pi)^{4(n-1)}f_1^\gimel(y)\cdots f_n^\gimel(y),
\end{equation}
which, however, is probably best regarded purely as an intermediary in calculations, insofar as it
introduces an equivalent of virtual particles in this formalism; in general, Eq. (\ref{XiGimel})
does not imply the existence of a function $\xi(;f_1,...,f_n)$, if $M(u)=0$ for some values of $u$.
In terms of $G$-modified functions, we can write equation (\ref{CommRel}) as
\vspace{-0.5ex}
\begin{equation}\label{gimelCommRel}
  [a^{\ }_{f^\gimel},a^\dagger_{g^\gimel}]=(2\pi)^4\left[(\!(g^\gimel;f^\gimel)\!)+
                          \lambda\left(a^{\;}_{g^{\gimel*}f^\gimel}+a^{\dagger}_{f^{\gimel*}g^\gimel}\right)\right],
\end{equation}
where $(\!(\cdot;\cdot)\!)$ is a ``plain'' inner product, with $M(u)=1$.

\section{Measurement and scattering}\label{MeasurementAndScattering}
For non-vacuum states, expressions for the classical $n$-measurement connected correlation functions can
be calculated easily enough, but do not as easily simplify.
For the state $\left|g\right>=\frac{\hat\phi_g}{\sqrt{(g;g)}}\left|0\right>$, for example, for $\hat\phi_f$,
$\hat\phi_f^2$, and $\hat\phi_f^3$, we have
\begin{widetext}
\begin{eqnarray}\label{Scattering}
  \left<g\right|\hat\phi_f\left|g\right>_{\hspace{0.3em}}
        &=&   \frac{2(g;g,f)}{(g;g)},\cr
  \left<g\right|\hat\phi_f^2\left|g\right>_c
        &=& (;f,f)+\frac{6(g;g,f,f)+2(g;f)(;g,f)}{(g;g)} - \frac{4(g;g,f)^2}{(g;g)^2},\cr
  \left<g\right|\hat\phi_f^3\left|g\right>_c
        &=& 2(;f,f,f) + \frac{6(g;f)(;g,f,f)+6(;g,f)(g;f,f)+24(g;g,f,f,f)}{(g;g)}\qquad\cr
        & &    \quad-\,\frac{12(g;g,f)(g;f)(;g,f)+36(g;g,f)(g;g,f,f)}{(g;g)^2} + \frac{16(g;g,f)^3}{(g;g)^3}.
\end{eqnarray}
\end{widetext}
A candidate form of causality for classical Lie fields is to require that states such as the above use only
functions that have positive spectrum in their construction.
If we could also apply this spectrum restriction to measurement operators, we would effectively recover the
spectrum condition of Wightman fields, since $G(u)$ always occurs together with test functions in $(...;...)$.
Local field operators cannot be required to satisfy this restriction, however, because it is incompatible with
reality, $\hat\phi_f^\dagger=\hat\phi_f$, which requires $f=f^*$.

Eigenvectors of space-time translations of the above \mbox{$\star$-algebra} can be constructed in a similar way to
those of free quantum fields.
We translate a state by translating the creation operators that were used to construct the state, so we introduce
a derivation on the \mbox{$\star$-algebra},
\begin{equation}
  \left[\partial_\alpha,a^\dagger_{g^*_y}\right]=a^\dagger_{\frac{\partial g^*_y}{\partial y^\alpha}},\qquad
         \partial_\alpha\left|0\right>=0,
\end{equation}
where $g_y(x)=g(x-y)$ is a space-time translation of the test function $g$.
We abbreviate this notation by assuming that all test functions are indexed by a space-time translation vector
$y^\alpha$, which we then omit.
The \mbox{$\star$-algebra} of observables defined by equation (\ref{CommRel}) and extended by $\partial_\alpha$ satisfies the
Jacobi identity.
As in the free quantum field case, the (improper, infinite norm) 1-particle eigenvectors of $\partial_\alpha$ are
$a^\dagger_{g^*_k}\left|0\right>$, where $g_k(x)=e^{ik_\alpha x^\alpha}$.
We can also construct $a^\dagger_{g^*_{k_1}}a^\dagger_{g^*_{k_2}}\cdots a^\dagger_{g^*_{k_n}}\left|0\right>$ as
eigenvectors (which are also improper and infinite norm) of $\partial_\alpha$.
The candidate form of causality suggested above also ensures that the Hilbert space generated by positive
spectrum test functions satisfies the spectrum condition.

We are now in a position to consider analogues of scattering theory in this formalism.
Scattering theory in the usual sense of ``in'' and ``out'' fields is not possible, because this paper has adopted,
in philosophers' terms, a block world formalism for models, in which a state that models (or, if we prefer,
approximately describes) the world is given for all time (which is why we used active translation instead of
state evolution to construct the operator $\partial_\alpha$ in the previous paragraph).
The 1-particle eigenvector $a^\dagger_{g^*_k}\left|0\right>$ is problematic to deal with because it does not have
finite norm, so we will use a function $g^{\;}_S$ that has bounded support in a region $S$ of momentum space, instead of
$g_k$, together with an improper measurement operator $\hat\phi_{f_k}$, where
$\tilde f_k(u)=\Half(\delta(u-k)+\delta(u+k))$.
For these functions, the 1-measurement correlation function is
\begin{widetext}
\begin{eqnarray}
  \left<g^{\;}_S\right|\hat\phi_{f_k}\left|g^{\;}_S\right>
        &=&   \frac{2(g^{\;}_S;g^{\;}_S,f)}{(g^{\;}_S;g^{\;}_S)},\cr
        &=&   2\lambda\frac{\int\delta(u_1-v_1-v_2)\Intd^4u_1 \tilde G^*(u_1)\tilde g^*_S(u_1)
                                                   \Intd^4v_1 \tilde G(v_1)\tilde g^{\;}_S(v_1)
                                                   \Intd^4v_2 \tilde G(v_2)\tilde f_k(v_2)  }
                           {\int\delta(u_1-v_1)\Intd^4u_1 \tilde G^*(u_1)\tilde g^*_S(u_1)
                                               \Intd^4v_1 \tilde G(v_1)\tilde g^{\;}_S(v_1)}\cr
        &=&    |\tilde G(k)|^2\frac{\xi(g^{\;}_S;g^{\;}_S)(-k)+\xi(g^{\;}_S;g^{\;}_S)(k)}{(g^{\;}_S;g^{\;}_S)}. 
\end{eqnarray}
\end{widetext}
When $k$ is a time-like 4-vector such that there is a $u$ for which $u$ and $u+k$ are both contained in the
bounded momentum-space region $S$, there is in general a non-zero expectation for the observable $\hat\phi_{f_k}$.
As we tune $S$ to be smaller and smaller, so that $a^\dagger_{g_S^*}\left|0\right>$ approaches a pure momentum state,
there is non-zero expectation for $\hat\phi_{f_k}$ only for very small $k$.

A different approach, which introduces nonlocal ``particle'' measurements, is to consider probabilities expressed as
inner products such as
\begin{widetext}
\begin{eqnarray}
  \frac{\left|\left<f_{S}\right|\left.g^{\;}_{S_1}\right>\right|^2}
       {\left<f_{S}\right|\left.f_{S}\right>\left<g^{\;}_{S_1}\right|\left.g^{\;}_{S_1}\right>}
    &=& \frac{|(f^{\;}_S;g^{\;}_{S_1})|^2}{(f^{\;}_S;f^{\;}_S)(g^{\;}_{S_1};g^{\;}_{S_1})},\cr
  \frac{\left|\left<f_{S}\right|\left.g^{\;}_{S_1},g^{\;}_{S_2}\right>\right|^2}
       {\left<f_{S}\right|\left.f_{S}\right>\left<g^{\;}_{S_1},g^{\;}_{S_2}\right|\left.g^{\;}_{S_1},g^{\;}_{S_2}\right>}
    &=& \frac{|(f^{\;}_S;g^{\;}_{S_1},g^{\;}_{S_2})|^2}
             {(f^{\;}_S;f^{\;}_S)\left[(g^{\;}_{S_1};g^{\;}_{S_1})(g^{\;}_{S_2};g^{\;}_{S_2})+
              (g^{\;}_{S_1};g^{\;}_{S_2})(g^{\;}_{S_2};g^{\;}_{S_1})+
             3(g^{\;}_{S_1},g^{\;}_{S_2};g^{\;}_{S_1},g^{\;}_{S_2})\right] },
\end{eqnarray}
\end{widetext}
where $f^{\;}_S$, $g^{\;}_{S_1}$, and $g^{\;}_{S_2}$ have bounded supports $S$, $S_1$, and $S_2$ in momentum space.
These can be understood as probabilities that (with normalization implicit) states prepared as
$\left|g^{\;}_{S_1}\right>$ and as $\left|g^{\;}_{S_1},g^{\;}_{S_2}\right>$ will, on measurement,
``look like'' $\left|f_{S}\right>$.
These probabilities will be non-zero, respectively, if $S\cap S_1\not=\emptyset$ (just as for a free field), or if
$S\cap (S_1\DP S_2)\not=\emptyset$, where $S_1\DP S_2=\left\{k:k=k_1+k_2; k_1\in S_1, k_2\in S_2\right\}$.
It is useful to undertake a Gram--schmidt orthogonalization of the state
$\left|g^{\;}_1,g^{\;}_2\right>=a^\dagger_{g^*_1}a^\dagger_{g^*_2}\left|0\right>$ relative to the 1-particle states
$\left|f\right>=a^\dagger_{f^*}\left|0\right>$, which gives what can more-or-less properly be called a 2-particle
state,
\begin{eqnarray}\label{OrthogonalState2}
  \left|g^{\;}_1,g^{\;}_2\right>_{\!2}
        &=&\left|g^{\;}_1,g^{\;}_2\right>-\left|\xi(;g^{\;}_1,g^{\;}_2)\right>\cr
        &=&\left(a^\dagger_{g^*_{S_1}}a^\dagger_{g^*_2}-
                 a^\dagger_{\xi^*(;g^{\;}_1,g^{\;}_2)}\right)\left|0\right>,
\end{eqnarray}
for which $\left<f\right|\left.g^{\;}_1,g^{\;}_2\right>_{\!2}=0$ for all functions $f$, so that there is zero
probability that $\left|g^{\;}_1,g^{\;}_2\right>_{\!2}$ will ``look like'' a 1-particle state.
On the basis of this Gram-Schmidt orthogonalization, $\left|g^{\;}_1,g^{\;}_2\right>$ should be understood to be a
superposition of a 1-particle state and a 2-particle state.
If an experiment prepares a state that looks sometimes like a 1-particle state and sometimes like a 2-particle state, then
the state should be modelled either as a superposition or as a mixture of 1- and 2-particle states.
Note that if $g_1$ and $g_2$ are improper restrictions to pure momenta $k_1$ and $k_2$, then both components of
$\left|g^{\;}_1,g^{\;}_2\right>_{\!2}$ are $(k_1+k_2)_\alpha$ eigenstates of $\partial_\alpha$.
For 3-particle states that are orthogonal to both 1-particle and 2-particle states, and for 4-particle states that are
orthogonal to 1-, 2-, and 3-particle states, we have
\begin{widetext}
\begin{eqnarray}
  \left|g^{\;}_1,g^{\;}_2,g^{\;}_3\right>_{\!3}&=&\left|g^{\;}_1,g^{\;}_2,g^{\;}_3\right>-
         \left|g^{\;}_1,\xi(;g^{\;}_2,g^{\;}_3)\right>_{\!2}-
         \left|g^{\;}_2,\xi(;g^{\;}_3,g^{\;}_1)\right>_{\!2}-
         \left|g^{\;}_3,\xi(;g^{\;}_1,g^{\;}_2)\right>_{\!2}-
         \left|\xi(;g^{\;}_1,g^{\;}_2,g^{\;}_3)\right>\cr
       &=&\left|g^{\;}_1,g^{\;}_2,g^{\;}_3\right>-
     \sum_\sigma\left[\SmallFrac{1}{2!}\left|g^{\;}_{\sigma(1)},\xi(;g^{\;}_{\sigma(2)},g^{\;}_{\sigma(3)})\right>_{\!2}+
                      \SmallFrac{1}{3!}\left|\xi(;g^{\;}_{\sigma(1)},g^{\;}_{\sigma(2)},g^{\;}_{\sigma(3)})\right>\right],
     \label{OrthogonalState3}\\
  \left|g^{\;}_1,g^{\;}_2,g^{\;}_3,g^{\;}_4\right>_{\!4}&=&
         \left|g^{\;}_1,g^{\;}_2,g^{\;}_3,g^{\;}_4\right>-
     \sum_\sigma\Bigl[
  \SmallFrac{1}{(2!)^2}\left|g^{\;}_{\sigma(1)},g^{\;}_{\sigma(2)},\xi(;g^{\;}_{\sigma(3)},g^{\;}_{\sigma(4)})\right>_{\!3}\!+
  \SmallFrac{1}{2!(2!)^2}\left|\xi(;g^{\;}_{\sigma(1)},g^{\;}_{\sigma(2)}),
                        \xi(;g^{\;}_{\sigma(3)},g^{\;}_{\sigma(4)})\right>_{\!2}\!+\cr
&&\qquad\hspace{7em}
  \SmallFrac{1}{3!}\left|g^{\;}_{\sigma(1)},\xi(;g^{\;}_{\sigma(2)},g^{\;}_{\sigma(3)},g^{\;}_{\sigma(4)})\right>_{\!2}+
  \SmallFrac{1}{4!}\left|\xi(;g^{\;}_{\sigma(1)},g^{\;}_{\sigma(2)},g^{\;}_{\sigma(3)},g^{\;}_{\sigma(4)})\right>
                   \Bigr].\label{OrthogonalState4}
\end{eqnarray}
If we consider the inner product between two 2-particle states,
\begin{equation}\label{IP2}
  \BLow{2}\left<g^{\;}_{S_3},g^{\;}_{S_4}\right|\left.g^{\;}_{S_1},g^{\;}_{S_2}\right>_{\!2}=
    (g^{\;}_{S_3};g^{\;}_{S_1})(g^{\;}_{S_4};g^{\;}_{S_2})+(g^{\;}_{S_4};g^{\;}_{S_1})(g^{\;}_{S_3};g^{\;}_{S_2})+
    2(g^{\;}_{S_3},g^{\;}_{S_4};g^{\;}_{S_1},g^{\;}_{S_2}),
\end{equation}
\end{widetext}
we find that there is a non-zero probability of $\left|g^{\;}_{S_1},g^{\;}_{S_2}\right>_{\!2}$ looking like
$\left|g^{\;}_{S_3},g^{\;}_{S_4}\right>_{\!2}$ if $S_3\cap S_1\not=\emptyset$ and $S_4\cap S_2\not=\emptyset$, or if
$S_4\cap S_1\not=\emptyset$ and $S_3\cap S_2\not=\emptyset$, which correspond to no interaction between the two particles,
or if $(S_3\DP S_4)\cap (S_1\DP S_2)\not=\emptyset$, which corresponds to an energy-momentum conserving interaction
between two particles, with the precise probabilities of observing a particular pair of particles instead of a different
pair being determined by the function $\tilde G(u)$.

Whereas the improper free field vectors $\left|K_1,K_2\right>$ and $\left|L_1,L_2\right>$ are orthogonal if the
wave numbers $K_1$ and $K_2$ are not a permutation of $L_1$ and $L_2$, we see from equation (\ref{IP2}) that the
classical random Lie field vectors $\left|K_1,K_2\right>_{\!2}$ and $\left|L_1,L_2\right>_{\!2}$ are orthogonal
only if, as well, $K_1+K_2\not=L_1+L_2$.
We can represent the inner product $\BLow{2}\left<K_1,K_2\right|\left.L_1,L_2\right>_{\!2}$ graphically as
\setlength{\unitlength}{2ex}
\begin{picture}(4.4,1.5)
  \multiput(0.1,0)(0.6,0){2}{\line(0,1){1.4}}
  \put(1.3,0.1){\makebox(1,1){$\,+\,2$}}
  \put(2.8,0){\line(1,1){1.4}}
  \put(4.2,0){\line(-1,1){1.4}}
\end{picture}.
Similarly, referring to equation (\ref{GSIP3}), $\left|K_1,K_2,K_3\right>_{\!3}$ and
$\left|L_1,L_2,L_3\right>_{\!3}$ are orthogonal if the wave numbers $K_1$, $K_2$, and $K_3$ are not a
permutation of $L_1$, $L_2$, and $L_3$, there are no permutations $\sigma,\tau\in S_3$ that result in
$K_{\sigma(1)}+K_{\sigma(2)}=L_{\tau(1)}+L_{\tau(2)}$ and $K_{\sigma(3)}=L_{\tau(3)}$, and
$K_1+K_2+K_3\not=L_1+L_2+L_3$, which we can represent graphically as
\begin{picture}(9.8,1.6)
  \multiput(0.1,0)(0.6,0){3}{\line(0,1){1.4}}
  \put(1.9,0.1){\makebox(1,1){$\,+\,2$}}
  \put(3.8,0){\line(0,1){1.4}}
  \put(4.1,0){\line(1,1){1.4}}
  \put(5.5,0){\line(-1,1){1.4}}
  \put(6.2,0.1){\makebox(1,1){$\,+\,12$}}
  \put(8.2,0){\line(1,1){1.4}}
  \put(8.9,0){\line(0,1){1.4}}
  \put(9.6,0){\line(-1,1){1.4}}  
\end{picture}.
Equation (\ref{GSIP4}) can be represented as\\
\centerline{
\begin{picture}(20.6,1.6)
  \multiput(0.1,0)(0.6,0){4}{\line(0,1){1.4}}
  \put(2.5,0.1){\makebox(1,1){$\,+\,2$}}
  \multiput(4.4,0)(0.6,0){2}{\line(0,1){1.4}}
  \put(5.3,0){\line(1,1){1.4}}
  \put(6.7,0){\line(-1,1){1.4}}
  \put(7.4,0.1){\makebox(1,1){$\,+\,12$}}
  \put(9.6,0){\line(0,1){1.4}}
  \put(9.9,0){\line(1,1){1.4}}
  \put(10.6,0){\line(0,1){1.4}}
  \put(11.3,0){\line(-1,1){1.4}}
  \put(11.4,0.1){\makebox(1,1){$\,+\,4$}}
  \multiput(13.0,0)(1.7,0){2}{\put(0,0){\line(1,1){1.4}}\put(1.4,0){\line(-1,1){1.4}} }  
  \put(16.8,0.1){\makebox(1,1){$\,+\,144$}}
  \put(19.0,0){\line(1,1){1.4}}
  \put(19.38,0){\line(1,2){0.7}}
  \put(20.02,0){\line(-1,2){0.7}}
  \put(20.4,0){\line(-1,1){1.4}}
\end{picture},}
and in general \emph{every} partition of $n$ contributes terms to
$\BLow{n}\left<K_1, ..., K_n\right|\left.L_1, ..., L_n\right>_{\!n}$.

The 4-dimensional block world context leads to significant differences from a conventional scattering theory,
nonetheless measurement operators such as $\left|f^{\;}_S\right>\left<f^{\;}_S\right|$
and $\left|g^{\;}_{S_1},g^{\;}_{S_2}\right>_{\!2}{\BLow{2}}\left<g^{\;}_{S_1},g^{\;}_{S_2}\right|$
generate a nonlocal, noncommutative \mbox{$\star$-algebra} that looks moderately familiar as a particle model.
If we restrict to positive frequency test functions in this nonlocal ``particle'' approach to measurement, we obtain a
formalism that is a quantum field theory, insofar as both state preparation and measurement use only positive
frequency test functions.
We cannot construct a \mbox{$\star$-algebra} of local observables that has non-trivially microcausal commutation
relations, but classical commutation relations between local field observables are possible.
An algebra of nonlocal ``particle'' observables such as $\left|f^{\;}_S\right>\left<f^{\;}_S\right|$ is in principle
non-observable and rather different from the commutative algebra generated by the local field observables
$\hat\phi_f$, with $f$ in the Schwartz function space, but particle observables are usable as an approximation.

An interesting mathematical problem is the extension of the above construction to limiting expressions which
introduce infinite products of the field, such as the characteristic function
$\left<0\right|e^{i\lambda\hat\phi_f}\left|0\right>$, which we can write formally as
$\exp{\!\left[\sum_{n=2}^\infty\frac{i^n\lambda^n}{n}(;f^{\times n})\right]}$, but with obvious worries
about convergence.
Although all correlation functions of finite degree are finite (which costs such effort to achieve in
conventional perturbation theory), $\xi(;f^{\times n})$ is a product of test functions in real space, which
is not in the Schwartz space for an infinite product of test functions.
The correlation functions of finite degree are in principle the empirically verifiable content of the
theory, however, so it seems easier to discount this mathematical problem than to discount the
mathematical problem of renormalization.

\section{Lie fields based on vector and electromagnetic fields}\label{nonScalarLieFields}
The dynamics of the quantized electromagnetic field in terms of a positive semi-definite inner product
on test functions is given by Menikoff and Sharp\cite[equation (3.27)]{MenikoffSharp}
(except for a missing factor of $(2\pi)^{-3}$ that is present in their equation (3.25)):
\begin{eqnarray}\label{EMInnerProduct}
     (E;F)_{EM} &=& \hbar\int\frac{\Intd^4u}{(2\pi)^4}
             2\pi\delta(u_\alpha u^\alpha)\theta(u_0)
             \tilde E_{\mu\beta}^*(u)u^\mu u^\nu\tilde F_{\nu}^{\ \beta}(u).\cr
&&
\end{eqnarray}
Note that $E$ and $F$ are \emph{not} electromagnetic field tensors, they are bivectors of
classical test functions that contribute to a description of measurement and/or state preparation
of the quantized electromagnetic field.
If the fourier transform $\tilde F_{\mu\nu}(u)$ of a test function $F$ has electric and magnetic field
components $(\tilde e_1,\tilde e_2,\tilde e_3)$ and $(\tilde b_1,\tilde b_2,\tilde b_3)$, the integrand
for the inner product $(F;F)_{EM}$ at $(u_0,0,0,u_0)$ is the positive semi-definite form
$u_0^2\left[(\tilde e_1+\tilde b_2)^2+(\tilde e_2-\tilde b_1)^2\right]$, which
suppresses all except two degrees of freedom of the quantum electromagnetic field at each wave number.
For a general free classical vector or bivector random field with nontrivial quantum fluctuations, in which we
remove the restriction to the forward light-cone found in equation (\ref{EMInnerProduct}), we
have available the inner products
\begin{widetext}
\begin{eqnarray}
     (U;V)_V &=& \hbar\int\Intd^4u\tilde U_\mu^*(u)
                  \left(\bigl(A_t(u)+A_s(u)\bigr)\frac{u^\mu u^\nu}{u^\alpha u_\alpha}-
                        A_s(u)g^{\mu\nu}\right)\tilde V^{\;}_\nu(u),\\
     (E;F)_B &=& \hbar\int\Intd^4u\Bigl[B(u)\tilde E_{\mu\beta}^*(u)u^\mu u^\nu\tilde F_{\nu}^{\ \beta}(u)+
        B_d(u)*\hspace{-0.65ex}\tilde E_{\mu\beta}^*(u)u^\mu u^\nu*\hspace{-0.65ex}\tilde F_{\nu}^{\ \beta}(u)\Bigr]
\end{eqnarray}
\end{widetext}
for positive scalar functions $A_t(u)$, $A_s(u)$, $B(u)$, and $B_d(u)$ (which in the free field case can be
distributions, concentrated on a single mass).
$*\hspace{-0.12ex}F_{\mu\nu}=\varepsilon_{\mu\nu}^{\hspace{1.0em}\alpha\beta}F_{\alpha\beta}$ is the Hodge dual of
the test function $F_{\alpha\beta}$.
In the vector case, $A_t(u)$ and $A_s(u)$ determine the independent amplitudes of fluctuations of, respectively,
time-like and space-like components of the vector field, and $A_t(u)=A_s(u)=0$ when $u^\alpha u_\alpha\le 0$; in
the bivector case, $B(u)$ and $B_d(u)$ determine the independent amplitudes of fluctuations of two sets of
Hodge dual components of the bivector field, and $B(u)=B_d(u)=0$ when $u^\alpha u_\alpha<0$.

For the construction of interacting creation and annihilation operators satisfying equation (\ref{CommRel}) as a
deformation of a free classical vector or bivector random field, we have to be able to construct generalized linear
forms $(...;...)$ and $\xi(...;...)$ that are symmetric in their anti-linear and linear parameters respectively,
and more generally that satisfy equations (\ref{C1})--(\ref{C3}) and (\ref{MCcondition}), which preclude any simple
construction of $\xi(...;...)$ using contraction directly.
We will therefore introduce a pair of Lorentz covariant sets of algebraic operators $q^{\alpha\dagger}$ and $q^\alpha$,
satisfying the commutation relation $[q^\alpha,q^{\beta\dagger}]=g^{\alpha\beta}$, and a pair of vectors
$\left|0_q\right>$ and $\left|0^\prime_q\right>$, for which $q^\alpha\left|0_q\right>=0$ and
$q^{\alpha\dagger}\left|0'_q\right>=0$ (the first of which yields $+g^{\mu\nu}$ at an appropriate point, the
second of which yields $-g^{\mu\nu}$).
With this definition, the operators $Q^\alpha=q^\alpha+q^{\alpha\dagger}=Q^{\alpha\dagger}$ generate a
commutative algebra, $[Q^\alpha,Q^{\beta}]=0$, so that we can contract test functions with $Q^\alpha$, then use
$\left<0_q\right|\cdots\left|0_q\right>$ to construct an inner product.
For $(U;V)_V$ and the electromagnetic half of $(E;F)_B$ (that is, for $B_d(u)=0$), we can write
\begin{eqnarray}
   (U;V)_V &=& \hbar\left<0'_q\right|\int\Intd^4u\tilde U^{\gimel*}(u) \tilde V^{\gimel}(u)\left|0'_q\right>,\\
   (E;F)_B &=& \hbar \left<0_q\right|\int\Intd^4u\tilde E^{\gimel*}(u) \tilde F^{\gimel}(u)\left|0_q\right>,
\end{eqnarray}
where 
\begin{eqnarray*}
  \tilde V^\gimel(u)&=&\tilde V_\mu(u)\left(\frac{\sqrt{A_t(u)+A_s(u)}}{\sqrt{u_\alpha u^\alpha}}u^\mu+
                                                                 \sqrt{A_s(u)}Q^\mu\right),\cr
  \tilde F^\gimel(u)&=&\tilde F_{\mu\nu}(u)\sqrt{B(u)}u^\mu Q^\nu;
\end{eqnarray*}
arbitrary complex phases may also be introduced into the definitions of $\tilde V^\gimel(u)$ and $\tilde F^\gimel(u)$.
The addition of the commutative algebra generated by $Q^\alpha$ leaves the derivations of section \ref{intKG}
essentially unchanged, so that we can ignore the vector and bivector indices, except when we consider final
expressions such as $(;F_1,F_2,F_3,F_4)$, for which we obtain
\begin{widetext}
\begin{equation}
  (;F_1,F_2,F_3,F_4)=(2\pi)^{12}\lambda^2\int\Intd^4y
         F^\gimel_{1\alpha_1}(y)F^\gimel_{2\alpha_2}(y)F^\gimel_{3\alpha_3}(y)F^\gimel_{4\alpha_4}(y)
         \left<0_q\right|Q^{\alpha_1}Q^{\alpha_2}Q^{\alpha_3}Q^{\alpha_4}\left|0_q\right>,
\end{equation}
\end{widetext}
where $\tilde F^\gimel_{1\alpha_1}(u)=\tilde F_{1\mu\alpha_1}(u)\sqrt{B(u)}u^\mu$, and 
$\left<0_q\right|Q^{\alpha_1}Q^{\alpha_2}Q^{\alpha_3}Q^{\alpha_4}\left|0_q\right>=
     g^{\alpha_1\alpha_2}g^{\alpha_3\alpha_4}+
     g^{\alpha_1\alpha_3}g^{\alpha_2\alpha_4}+
     g^{\alpha_1\alpha_4}g^{\alpha_2\alpha_3}$
is a symmetric sum of products of two metric tensors.
We could define $(;F_1,F_2,F_3,F_4)$ directly as a symmetrized contraction, but an algebraic construction
emphasizes that the algebraic and combinatoric structures of scalar, vector, and bivector fields are identical,
up to the final step of evaluating inner products and terms such as $(;F_1,F_2,F_3,F_4)$ as space-time integrals.
The introduction of the algebra generated by $Q^\alpha$ effectively extends the test function
space to the symmetrized tensor algebra of the tangent space over space-time, since we have implicitly introduced
operators such as $a^\dagger_{F^\gimel_{1\alpha_1}F^\gimel_{2\alpha_2}F^\gimel_{3\alpha_3}F^\gimel_{4\alpha_4}
         Q^{\alpha_1}Q^{\alpha_2}Q^{\alpha_3}Q^{\alpha_4}}$; it is possible, however, for us to take the extended
test function space to be purely internal, analogously to the $\xi^\gimel$ of Eq. (\ref{XiGimel}).

\appendix
\section{Measurement theory for quantum fields and for classical random fields}\label{MeasurementTheory}
Conventionally, quantum mechanics models a set of measurements in an idealized way as an incompatible set of
operators acting on a single state in a Hilbert space, which represents a measured system, but for more detailed
models the measurement apparatus must be included.
For example, in Feynman and Hibbs,
\emph{``The usual separation of observer and observed which is now needed in analyzing measurements in quantum
mechanics should not really be necessary, or at least should be even more thoroughly analyzed. What seems
to be needed is the statistical mechanics of amplifying apparatus''}\cite[pp. 22-23]{FeynmanHibbs}.
More forcefully, Bell urges us to describe experiments instead of speaking of measurements of systems\cite{BellAM}.
Once preparation and measurement apparatuses are included in a detailed model of an experiment, the difference between
quantum field models and classical random field models for experiments becomes nominal and empirically irrelevant.
What must be particularly ensured, however, is that the statistical mechanics of amplifying apparatus includes the effects
of quantum fluctuations, either implicitly in a quantum measurement model or explicitly in a classical measurement
model, whenever considering the effects of quantum fluctuations is necessary to ensure empirical adequacy of a model.

It should be noted that classical random fields are sufficiently similar to quantum fields and sufficiently different
from classical particle property models that the assumptions that are needed to derive Bell inequalities are generally
not satisfied if there are any thermal or quantum fluctuations\cite{MorganC}.

It is a curiosity of the measurement theory of quantum fields that modelling experiments using measurement operators
places measurement apparatuses and the Physicist outside of space and time, and supposes that the use of different
measurement apparatuses \emph{in principle} does not change the physical state of any preparation apparatus (that is
included in the model) or of the measured system \emph{whatsoever}.
In contrast, if we include the measurement apparatus in a quantum field model, the presence of the measurement
apparatus certainly changes the measured system -- the past of the measurement apparatus is in the past light-cone
of the measured system.
Incompatibility of measurement operators can be thought of loosely as describing the non-trivial effects of
measurements on each other, while curiously requiring the measured state in principle not to be different when
we make the different measurements.
To determine a quantum state, however, we \emph{have} to make incompatible measurements of the \emph{same} prepared state.
This is perfectly acceptable as a pragmatic approach to measurement, but, as a result, moving the Heisenberg cut changes
the relationship between measurement apparatus and measured system, so that the measured system either is or is not
changed by the presence of the measurement apparatus.

If we consider classical random fields in terms of measurement, then they have their own, equally pragmatic assumption
about measurement: that we can make measurements that effectively do not change the measured system.
This assumption is violated quite strongly when we attempt, for example, to describe electromagnetic fields
in the quantum mechanical regime as classical systems, insofar as from a classical point of view the measurement
apparatus does change the measured system.
Instead of considering classical random fields in terms of measurement, however, it is conceptually more straightforward
to think of classical random fields as allowing us to construct \emph{models}, or descriptions, of experiments.
The models we construct are idealized to a greater or lesser extent, depending on how much care is needed to
achieve sufficient empirical accuracy.
If quantum fluctuations are empirically significant, which they are when we describe experiments that are in
the quantum mechanical regime using classical random fields, we have to include in the model the effects of
quantum fluctuations of all interacting parts of the experiment.
In practice, this presumably never includes the person who is watching a computer display of experimental results, but
it often will require us to take account of the effect of quantum fluctuations on interactions between measurement
apparatus and the electromagnetic field, ``the statistical mechanics of amplifying apparatus''.

In reverting to a classical random field model that explicitly includes experimental apparatuses and the
effects of quantum fluctuations as well as the effects of thermal fluctuations in models, we treat thermal
and quantum fluctuations in a more even-handed way.
If we understand the Unruh effect to describe transformations between quantum and thermal fluctuations under
non-inertial changes of coordinates, the principle of equivalence and a preference for covariant presentation
of physical models suggest that we ought to prefer classical random field models to quantum field models.
Making the expressions given in this paper for $(...;...)$ and $\xi(...;...)$ generally covariant and independent of or
determining the underlying geometry, however, is a nontrivial problem.

One situation in which quantum fluctuations are physically significant but not necessarily a significant
part of measurement, so that classical random field models might be as effective as quantum field
models, is in solid state physics, insofar as quantum fluctuations are important to the internal structure
of a material but measurement apparatuses measure thermodynamical properties without quantum fluctuations
of the measurement apparatus significantly affecting the microstructure of the material.
It also seems that with the addition of detailed consideration of the quantum fluctuations of light sources,
detectors, and the electromagnetic field, we might reasonably expect to expand the scope of semi-classical optics,
which is already empirically adequate for a wide range of experiments.
It is always a contingent question whether a detailed model of sources and detectors is necessary
to model a whole experiment with sufficient empirical adequacy, whether we model experiments
in terms of a quantum mechanical or a classical model of measurement.
The advantage of a quantum mechanical model of measurement is of course that it is often not necessary
to model measurement apparatuses explicitly, where a classical model of measurement for the same experiments
may require detailed explicit models of quantum fluctuations.

\vspace{3ex plus 5ex}

\section{The vacuum state of the classical random Lie field}\label{VacuumState}
To show that $\left<0\right|A\left|0\right>$ is a state over the classical random Lie field \mbox{$\star$-algebra},
we have to show that $\left<0\right|AA^\dagger\left|0\right>$ is positive semi-definite
for operators $A$ constructed only from annihilation operators.
For a classical random Lie field, the multilinear form $\xi(;f_1,f_2,...,f_n)$ can be defined to be
$\xi(f_1^*;f_2,...,f_n)$, which we can use to construct the orthogonal states $\left|g_1,g_2\right>_{\!2}$,
$\left|g_1,g_2,g_3\right>_{\!3}$, and $\left|g_1,g_2,g_3,g_4\right>_{\!4}$
of equations (\ref{OrthogonalState2}), (\ref{OrthogonalState3}), and (\ref{OrthogonalState4}).
Computer algebra verifies fairly straightforwardly that these states, and the similarly defined state
$\left|g_1,g_2,g_3,g_4,g_5\right>_{\!5}$, are mutually orthogonal for all choices of functions, with
inner product a sum over partitions of $n$ and twice over permutations in $S_n$,
\begin{widetext}
\begin{equation}\label{GSIP}
  \BLow{m}\left<f_1,...,f_m\right|\left.g_1,...,g_n\right>_{\!n}=
       \delta_{m,n}\sum_\pi \frac{M_2(\pi)}{n!}\sum_{\sigma\in S_n}\sum_{\tau\in S_n}
           P_\pi(f_{\sigma(1)},...f_{\sigma(n)};g_{\tau(1)},...,g_{\tau(n)})
\end{equation}
where $M_2(\pi)=n!/1^{a_1}a_1!\,2^{a_2}a_2!\,...n^{a_n}a_n!$ (with $a_k$ the number of times that $k$ appears in
the partition $\pi$, see \cite[(p823,\,p831)]{AbramowitzStegun}), and $P_{(a_1,a_2,...,a_n)}(f_1,...f_n;g_1,...,g_n)$
constructs a term with $\sum a_k$ factors, each of the form $(f_j,...,f_{j+k-1};g_j,...,g_{j+k-1})$; for example,
\begin{equation}\label{P21}
  P_{(2,1)}(f_1,f_2,f_3,f_4;g_1,g_2,g_3,g_4)=(f_1;g_1)(f_2;g_2)(f_3,f_4;g_3,g_4),
\end{equation}
\end{widetext}
where $(2,1)$ is an unambiguous abbreviation of $(2,1,0,0)$.
A verification that Eq. (\ref{GSIP}) is correct for the first five of the states $\left|g_1,...,g_n\right>$
is of course not a proof, which I cannot currently provide, however the very straightforward algebraic and
combinatorial structure strongly suggests that it is a correct conjecture for all $m,n$.
Including multiplicity, $\BLow{n}\left<f_1,...,f_m\right|\left.g_1,...,g_n\right>_{\!n}$ generates
$n!^2$ terms, in contrast to $n!$ terms for the free field (which correspond to the partition $(n)$ in equation
(\ref{GSIP})).
Computer algebra beyond
$\BLow{5}\left<f_1,f_2,f_3,f_4,f_5\right|\left.g_1,g_2,g_3,g_4,g_5\right>_{\!5}$
requires more computing resources than I have available (more efficient programming would probably
extend my limit to 6, but the improvement would only be cosmetic).
The first five inner products are:
\begin{widetext}
\begin{eqnarray}
  \left<f_1\right|\left.g_1\right>&=&(f_1;g_1),\\
  \BLow{2}\left<f_1,f_2\right|\left.g_1,g_2\right>_{\!2}&=&(f_1;g_1)(f_2;g_2)+(f_2;g_1)(f_1;g_2)+
                                                            2(f_1,f_2;g_1,g_2),\label{GSIP2}\\
  \BLow{3}\left<f_1,f_2,f_3\right|\left.g_1,g_2,g_3\right>_{\!3}&=&
         (f_1;g_1)(f_2;g_2)(f_3;g_3)+(f_1;g_1)(f_3;g_2)(f_2;g_3)+(f_2;g_2)(f_1;g_3)(f_3;g_1)+\cr
&&\quad  (f_2;g_1)(f_1;g_2)(f_3;g_3)+(f_2;g_1)(f_1;g_3)(f_3;g_2)+(f_1;g_2)(f_2;g_3)(f_3;g_1)+\cr
&&\quad 2(f_1;g_1)(f_2,f_3;g_2,g_3)+2(f_1;g_2)(f_2,f_3;g_1,g_3)+2(f_1;g_3)(f_2,f_3;g_1,g_2)+\cr
&&\quad 2(f_2;g_1)(f_1,f_3;g_2,g_3)+2(f_2;g_2)(f_1,f_3;g_1,g_3)+2(f_2;g_3)(f_1,f_3;g_1,g_2)+\cr
&&\quad 2(f_3;g_1)(f_1,f_2;g_2,g_3)+2(f_3;g_2)(f_1,f_2;g_1,g_3)+2(f_3;g_3)(f_1,f_2;g_1,g_2)+\cr
&&\quad 12(f_1,f_2,f_3;g_1,g_2,g_3),\label{GSIP3}\\
  \BLow{4}\left<f_1,f_2,f_3,f_4\right|\left.g_1,g_2,g_3,g_4\right>_{\!4}&=&
          (f_1;g_1)(f_2;g_2)(f_3;g_3)(f_4;g_4)+ ... \qquad\qquad\qquad\, (24\mathrm{\ terms})\cr
&&\quad  +\;2(f_1;g_1)(f_2;g_2)(f_3,f_4;g_3,g_4)+ ... \qquad\qquad (72\mathrm{\ terms})\cr
&&\quad +\;4(f_1,f_2;g_1,g_2)(f_3,f_4;g_3,g_4)+ ... \qquad\qquad\,\, (18\mathrm{\ terms})\cr
&&\quad +\;12(f_1;g_1)(f_2,f_3,f_4;g_2,g_3,g_4)+ ... \qquad\qquad (16\mathrm{\ terms})\cr
&&\quad +\;144(f_1,f_2,f_3,f_4;g_1,g_2,g_3,g_4),\label{GSIP4}\\
  \BLow{5}\left<f_1,f_2,f_3,f_4,f_5\right|\left.g_1,g_2,g_3,g_4,g_5\right>_{\!5}&=&
          (f_1;g_1)(f_2;g_2)(f_3;g_3)(f_4;g_4)(f_5;g_5)+ ... \qquad\qquad\qquad\, (120\mathrm{\ terms})\cr
&&\quad  +\;2(f_1;g_1)(f_2;g_2)(f_3;g_3)(f_4,f_5;g_4,g_5)+ ... \qquad\qquad (600\mathrm{\ terms})\cr
&&\quad +\;4(f_1;g_1)(f_2,f_3;g_2,g_3)(f_4,f_5;g_4,g_5)+ ... \qquad\qquad\,\, (450\mathrm{\ terms})\cr
&&\quad +\;12(f_1;g_1)(f_2;g_2)(f_3,f_4,f_5;g_3,g_4,g_5)+ ... \qquad\qquad (200\mathrm{\ terms})\cr
&&\quad +\;24(f_1,f_2;g_1,g_2)(f_3,f_4,f_5;g_3,g_4,g_5)+ ... \qquad\qquad\,\, (100\mathrm{\ terms})\cr
&&\quad +\;144(f_1;g_1)(f_2,f_3,f_4,f_5;g_2,g_3,g_4,g_5)+ ... \qquad\qquad (25\mathrm{\ terms})\cr
&&\quad +\;2880(f_1,f_2,f_3,f_4,f_5;g_1,g_2,g_3,g_4,g_5).\label{GSIP5}
\end{eqnarray}
The number of permutations associated with each partition is $M_1(\pi)M_3(\pi)$, where
\begin{equation}
M_1(\pi)=n!/(1!)^{a_1}\,(2!)^{a_2}\,...\,(n!)^{a_n},    \qquad
M_3(\pi)=n!/(1!)^{a_1}a_1!\,(2!)^{a_2}a_2!\,...\,(n!)^{a_n}a_n!. 
\end{equation}
We will show that the sum over permutations,
\begin{equation}
  \mathbf{S}_\pi(f_1,...,f_n;g_1,...,g_n)=\sum_{\sigma\in S_n}\sum_{\tau\in S_n}
           P_\pi(f_{\sigma(1)},...f_{\sigma(n)};g_{\tau(1)},...,g_{\tau(n)})
\end{equation}
\end{widetext}
is an inner product for each partition independently.
Equation (\ref{GSIP2}) is the sum of the free field inner product (the Permanent of a Gram matrix $(f_i,g_j)$,
which is well-known to be an inner product over a symmetrized tensor product space\cite[\S 2.2]{Minc}) and
$2(f_1,f_2;g_1,g_2)$, which is also clearly an inner product.
In equation (\ref{GSIP3}), the first six terms are the free field inner product, and the last term is again
clearly an inner product.
The terms $(f_1;g_1)(f_2,f_3;g_2,g_3),...$ can be written as a trace of two matrices,
$\mathrm{Tr}\left[M(f,g)N(f,g)^T\right]$, where $M_{ij}=(f_i;g_j)$, $N_{ij}=(F_i;G_j)$;
$F_1=(f_2,f_3)$, etc.
$M$ and $N$ are both Gram matrices, so the trace $\mathrm{Tr}\left[M(f,f)N(f,f)^T\right]$ is
positive semi-definite for any decomposable symmetrized tensor $f_1\otimes f_2\otimes f_3$, hence
$\BLow{3}\left<f_1,f_2,f_3\right|\left.g_1,g_2,g_3\right>_{\!3}$ is also an inner product over the 
symmetrized tensor product space.
More generally, proceeding by induction,
\begin{equation}
\sum_{\sigma\in S_n}\sum_{\tau\in S_n}(f_{\sigma(1)};g_{\tau(1)})
           P_{\pi_1}(f_{\sigma(2)},...,f_{\sigma(n)};g_{\tau(2)},...,g_{\tau(n)})
\end{equation}
is $\mathrm{Tr}\left[M(f,g)N_{\pi_1}(f,g)^T\right]$, where $\pi_1$ is a partition of $n-1$ and
$N_{\pi_1}(f,g)$ is a Gram matrix of $n$ vectors $(f_1,...,\Remove{f_j},...,f_n)$ and $n$ vectors
$(g_1,...,\Remove{g_j},...,g_n)$ using $\mathbf{S}_{\pi_1}$, which, on the induction hypothesis, is an
inner product.
With the same construction, using $(f_{\sigma(1)},f_{\sigma(2)};g_{\tau(1)},g_{\tau(2)})$ \emph{etc.}
instead of $(f_{\sigma(1)};g_{\tau(1)})$, we can confirm by induction that $\mathbf{S}_\pi$ is an
inner product for each partition of $n$, and hence that $\left<0\right|A\left|0\right>$ is a state
over the classical random Lie field.

\end{document}